\documentclass{article}
\usepackage{glas}
\usepackage{graphicx}
\graphicspath{{eps/}{pics/}{figs/}}
\begin{document}
\begin{titlepage}{GLAS-PPE/97--05}{27 August 1997}

\title{Recent results on GaAs detectors - 137}

\author{R.L.~Bates\InstAnotref{glas}{\dagger}
M.Campbell\Instref{CERN}
C.~Da'Via\Instref{glas}
S.~D'Auria\Instref{udine}
V.~O'Shea\Instref{glas}
C.~Raine\Instref{glas}
K.M.~Smith\Instref{glas}}
\Instfoot{glas}{Dept. of Physics \& Astronomy, University of Glasgow, UK}
\Instfoot{CERN}{ECP Devision CERN, Geneva, Switzerland}
\Instfoot{udine} {Dip. di Fisica, Universit\`a di Udine, Italy}
\Anotfoot{\dagger}{Partially supported by a CASE award from the Rutherford Appleton Lab , UK}

\collaboration{On Behalf of the RD8 Collaboration}

\begin{abstract}
The present understanding of the charge collection in GaAs detectors with respect to the materials used and its processing are discussed. The radiation induced degradation of the charge collection efficiency and the leakage current of the detectors are summarised. 
The status of strip and pixel detectors for the ATLAS experiment are reported along with the latest results from GaAs X-ray detectors for non-high energy physics applications.
\end{abstract}
\end{titlepage}

\section{Introduction}

The purpose of the RD8 program \cite{RD8} is to develop radiation hard strip and pixel detectors from GaAs material as part of the research and development projects initiated by CERN for the LHC.
This paper outlines the latest results with regard to the understanding of the operation of GaAs detectors, their radiation hardness, and the present level of development of ATLAS design strip and pixel detectors. The pixel detectors have been tested within the RD19 collaboration \cite{RD8} using the latest LHC pixel readout chip. 

\section{Understanding GaAs Detectors}

To fabricate a GaAs detector two choices have to be made, namely the substrate material and the contacts used. Care must also be taken over surface preparation and passivation.

The GaAs used almost universally is semi-insulating undoped, SI-U, LEC industrial substrate material supplied by various manufacturers. Material with a low carbon concentration and chromium- or iron-doped material, supplied by SITP, Tomsk, have been proposed as alternatives for improved radiation hardness. Charge carrier absorption lengths vary from 100$\mu$m to 1600$\mu$m for holes and 100$\mu$m to 500$\mu$m for electrons \cite{aachen}. The low carbon material has the largest values for electrons. The spread in these values, even from a single manufacturer, is also considerable. For a 200$\mu$m thick detector it is desirable to have absorption lengths for both carriers in excess of 200$\mu$m.
The detector may be fabricated with either a p-i-n structure \cite{spyros} by doping the substrate material, or with a Schottky and an ohmic contact realized by metal deposition.

\begin{figure}
\centering
\resizebox{.7\textwidth}{!}{\includegraphics{{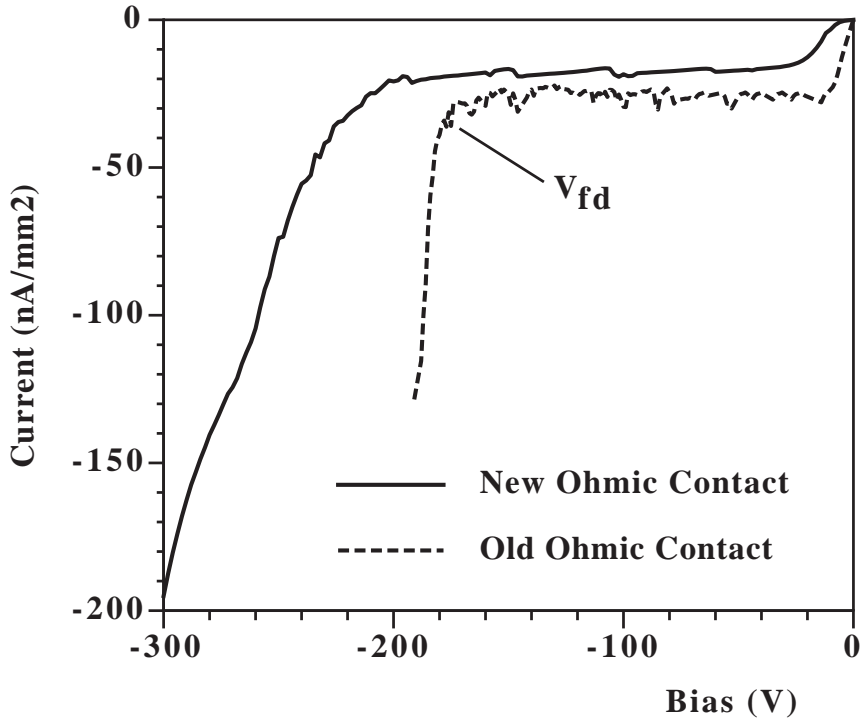}}}
\caption{The current voltage characteristics of two 200$\mu$m thick SI-U LEC GaAs diodes}\label{fig:iv}
\end{figure}

The leakage current of a typical LEC diode as a function of reverse bias is shown in figure \ref{fig:iv}. The value of the plateau current, which is independent of diode thickness, is between 10 and 30nA~mm$^{-2}$ at 20$^{o}$C. When a voltage corresponding to approximately one volt per micron of substrate thickness, V$_{fd}$, is reached the leakage current increases dramatically due to current injection through the ohmic contact. 
Also shown in the figure is the current characteristic of a diode with an improved ohmic contact where the leakage current increases only slowly above the bias V$_{fd}$. This reduced current at high voltages allows the detector to be operated up to voltages approaching twice V$_{fd}$. The improved ohmic contact was realised by annealing a multi-layer titanium-palladium-germanium metal contact.
Alenia SpA have fabricated diodes with an ion-implanted ohmic contact which allow a bias many times V$_{fd}$ to be applied \cite{nava2}.

\begin{figure}
\centering
\resizebox{.7\textwidth}{!}{\includegraphics{{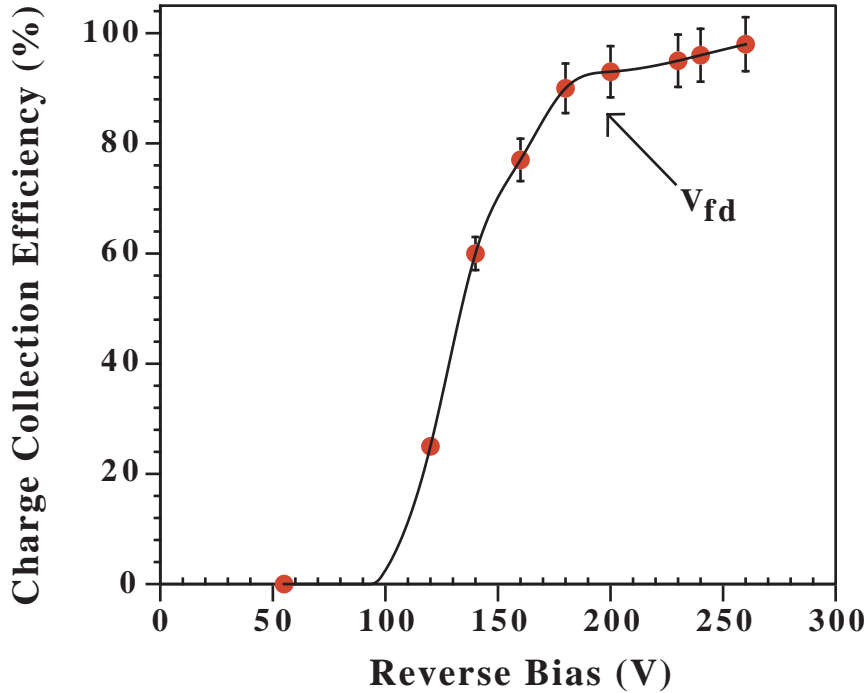}}}
\caption{The charge collection efficiency as a function of bias for a 200$\mu$m thick SI-U LEC GaAs pad detector}\label{fig:cce}
\end{figure}

The surface quality and thus preparation of the wafer has proved important in obtaining high charge collection efficiencies (CCE). The CCE increases linearly with applied bias (shown in figure \ref{fig:cce} for a typical detector) until the bias V$_{fd}$ is reached, where the efficiency for a 200$\mu$m detector is typically between 60\% and 80\%. At higher biases the efficiency slowly increases towards 100\%. 

Although high resistivity GaAs would be expected to deplete at a few volts this is not found to be the case. With an alpha particle source on the back, ohmic contact, for example, signals are only observed at bias voltages greater than V$_{fd}$.

The electric field distribution in the material consists of two regions: a high field region with an approximately constant value of 1~V/$\mu$m where charge collection is high and a low field region where almost no collection occurs. The penetration depth of the high field region has been measured at 0.7 to 1.0 $\mu$m per volt, depending upon material \cite{field1,field2,field3}. The field distribution explains the increase in leakage current at V$_{fd}$, the linear dependence of CCE on bias, and the observation of rear alpha signals only at biases in excess of V$_{fd}$.

\section{ATLAS detectors} 
\subsection{Strip Detectors}
 Last year GaAs microstrip detectors with parallel strip patterned Schottky contacts were tested at CERN \cite{aachen,man,thilo}. Different strip pitch and metal contact width-to-pitch ratios were used to investigate the effects of charge sharing and possible signal loss between strips. These detectors met LHC performance specifications with respect to signal to noise, resolution, and speed \cite{specs}.

The first wedge-shaped ATLAS prototype detectors from EEV and Alenia were tested at CERN in the H8 test beam this year. These have 65mm long Schottky fan-shaped strips with the pitch varying from 50$\mu$m at the narrow end to 76$\mu$m. The detectors have integrated bias structures and silicon nitride decoupling capacitors. The detectors were read out by either analogue (Felix \cite{felix}) or binary \cite{binary} electronics. One detector irradiated to a fluence of 7.7~10$^{13}$~24GeV/c~p/cm$^{2}$ was tested at room temperature with Felix readout. Two parallel strip detectors with integrated resistors and decoupling capacitors fabricated upon Tomsk substrates, one of which irradiated to a fluence of 3 10$^{13}$ p/cm$^{2}$, were also tested with Felix readout. Analysis of these data has just begun.

\subsection{Pixel Detectors}

The work reported in this section has been performed by members of RD8 and RD19.

\begin{figure}
\centering
\resizebox{.7\textwidth}{!}{\includegraphics{{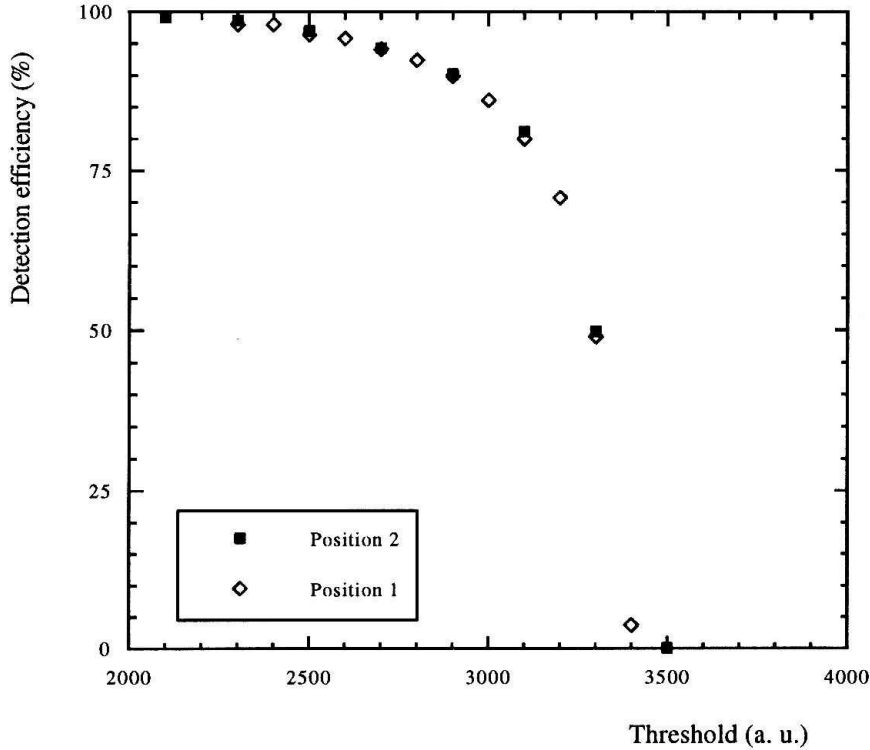}}}
\caption{Detection efficiency of GaAs pixel detector AL-IOC-10 with an ion-implanted ohmic contact as a function of Omega3 read-out chip comparator threshold value}\label{fig:pixels}
\end{figure}

GaAs pixel detectors have been developed which are compatible with the RD19 Omega3/LHC1 \cite{omega3} read-out electronics. The pixel detectors form an array of 16 x 128 elements each 50$\mu$m by 500$\mu$m in size. The latest pixel detectors fabricated by Alenia with an ion-implanted ohmic back contact have been tested in a 120GeV/c pion beam. The detectors were 200$\mu$m thick and biased to 500V. The detection efficiency dependence upon the Omega3 comparator threshold for the two planes of GaAs pixels under test in an Omega2 silicon telescope is shown in figure \ref{fig:pixels}. The pixels performed as well as the silicon telescope and demonstrated 100\% CCE. These results are from on-line analysis but show that GaAs pixels work extremely well.

\section{Radiation Hardness}

In the GaAs forward tracker detector of the ATLAS experiment the fluence is dominated by pions at an estimated level of 1.2 10$^{14}$ $\pi$/cm$^{2}$ over the experiment`s ten year lifetime \cite{garfine}. The neutron fluence is expected to be 3 to 10 times less that this, while that due to the other charged hadrons is small in comparison. The effects of particular radiation have therefore been investigated. Electrons and gamma rays have shown negligible effects. 
The increase in leakage current due to neutrons (n), protons (p), and pions ($\pi$) is small, typically at the 50\% level after 10$^{14}$ particles/cm$^{2}$, and the current exhibits a slow decrease with time at room temperature after irradiation. The value of V$_{fd}$ is typically half its pre-irradiation value after 3 10$^{13}$ p/cm$^{2}$ and remains unchanged at higher fluences. 

\begin{figure}
\centering
\resizebox{.7\textwidth}{!}{\includegraphics{{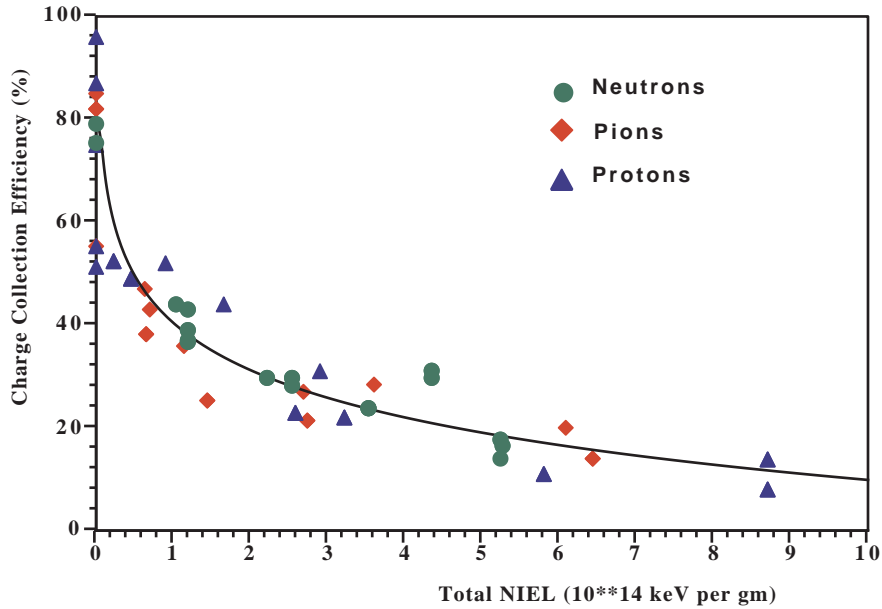}}}
\caption{The charge collection efficiency as a function of total NIEL for ISIS neutrons, 300MeV/c pions and 24GeV/c protons for 200$\mu$m thick SI-U LEC GaAs detectors}\label{fig:rad2}
\end{figure}

The charge collection, however, falls with fluence, with the hole signal component being more affected. The reduction of CCE with fluence depends on the type and energy of the irradiating particle. For a 200$\mu$m thick LEC detector the CCE falls to 20\% after 4 10$^{14}$ n/cm$^{2}$ (5300 electrons) or 1.5 10$^{14}$ p/cm$^{2}$ or only 6 10$^{13}$ pions/cm$^{2}$ \cite{meatworkshop}. The differences have been explained in term of the different non-ionising energy loss, NIEL, of the particles in the detector material, as shown in figure \ref{fig:rad2} \cite{terry}. 

All LEC material, except for the low carbon material, appears to degrade to give a minimum ionising particle signal of approximately 4500 electrons after a fluence of 1.6 10$^{14}$ p/cm$^{2}$. The latter material may be slightly more radiation hard due to the high initial values of the charge carrier mean free drift lengths and gives a signal of 7000 electrons after the same fluence \cite{xiao}. 

Diodes fabricated on Tomsk material have shown better radiation hardness \cite{tomsk}, with a reduction of 50\% in CCE after a fluence of 1.1 10$^{13}$ (1GeV/c) p/cm$^{2}$ at a flux of 5 10$^{14}$ p/hour. A dependence of the degradation of CCE on the rate of irradiation has been reported. With a flux of only 5 10$^{13}$ p/hour the 50\% CCE reduction occurs after a fluence of 1.5 10$^{14}$ p/cm$^{2}$. Proton irradiations have been performed between -5$^{o}$C and -10$^{o}$C on LEC material. Preliminary results suggest that the reduction in CCE is slightly worse than with room temperature irradiations.

\section{X-ray Detection}

A thickness of 200$\mu$m of GaAs has more than an order of magnitude higher X-ray absorption efficiency with respect to 300$\mu$m of silicon over the medically interesting energy range of 10 to 100keV. The detection properties of GaAs and Ge are similar but GaAs detectors can be operated at or just below room temperature. The increased absorption efficiency implies that a lower dose is required for imaging applications to produce an image of similar quality to that from film or silicon detectors.
A 100$\mu$m thick GaAs pad detector with an ion-implanted ohmic contact, biased at 500V, has been used to produce Am-241 X-ray spectra \cite{nava}. At 20$^{o}$C the energy resolution of 20\% allowed lines down to 20keV to be distinguished. At -30$^{o}$C the leakage current reduction allowed lines as low as 8keV to be resolved, with a resolution of 4\%.

Low pressure growth of VPE GaAs is being investigated as a possible source of affordable, high quality material for X-ray imaging. The first samples have recently been supplied for evaluation by Epitronics Corp. \cite{epi}. These are 80$\mu$m thick, with charge carrier concentration $\sim$ 2-3 10$^{14}$ /cm${^3}$. The material was found to deplete as expected from standard theory to a depth of 20$\mu$m for an applied bias of 100V. The CCE of simple pad detectors fabricated with this material was 100\% in the depleted region, allowing the Barium (32keV) K X-ray line to be resolved. The limiting factor was again the leakage current noise which was very high due to the rough surface of the VPE material. Careful mechanical and chemical polishing of the wafer can reduce the leakage current and thus allow better resolution to be obtained. 

GaAs pixel detectors bump-bonded to Omega3 read-out electronics have been used to image medical X-ray phantoms with Am-241 and Cd-109 X-ray sources \cite{davia}. The results are encouraging, with high signal-to-noise and signal-to-contrast ratios being obtained. The results were improved by the partial removal of Compton interactions using the fast-OR trigger of the electronics. Auto-radiography applications have been investigated by imaging beta particles from P-32 labelled human mammary cells. A low noise level of 10$^{-4}$~cps~mm$^{-2}$ and linearity for activities between 0.005nCi and 0.5nCi were obtained.  
 Further improvements are possible in the quality of both the VPE and LEC material and should allow better results to be obtained.

\section{Conclusions}

The extended knowledge of the mechanism of charge transport in GaAs has led to detectors with 100\% CCE. ATLAS prototype strip detectors have been fabricated and test-beam results are being analysed. The required specifications for ATLAS pixel detectors have been met. While the reduction in CCE due to charged hadron irradiation is severe in conventional LEC material, low carbon material or Tomsk technology may provide better radiation resistance. The flux dependence of CCE degradation has to be investigated further. X-ray detection with bulk-grown and VPE GaAs is under investigation. At present, leakage current noise limits the obtainable resolution for room temperature operation.

\section{Acknowledgements}

The authors would like to thank F.McDevitt, A.Meikle, F.Doherty and R.Boulter for technical support and all those at the ISIS, CERN PS and PSI facilities during the irradiation runs. One of us (R. Bates) gratefully acknowledges the support received through a CASE postgraduate student ship from RAL. The work carried out with pixel detectors could not have occurred without the collaboration of RD19 with RD8. The results obtained within the RD8 and RD19 collaborations are from work partly funded by PPARC (UK), INFN (Italy), and the BMFD (Germany).

\end{document}